\documentclass[12pt, a4paper]{article}
\usepackage{a4wide,amssymb,amsmath,amscd,}

\newcommand{\T}{{\mathbb T}}
\newcommand{\Z}{{\mathbb Z}}
\newcommand{\R}{{\mathbb R}}
\newcommand{\C}{{\mathbb C}}

\begin{document}

\topmargin -2pt

\headheight 0pt

\topskip 0mm \addtolength{\baselineskip}{0.20\baselineskip}
\begin{flushright}
{\tt math-ph/0501007} \\
{\tt KIAS-P04054}
\end{flushright}


\begin{center}

{\Large \bf  Symmetry of Quantum Torus  \\
with Crossed Product Algebra}\\

\vspace{10mm}

{\sc Ee Chang-Young}\footnote{cylee@sejong.ac.kr}\\
{\it Department of Physics, Sejong University, Seoul 143-747, Korea}\\


and \\


{\sc Hoil Kim}\footnote{hikim@knu.ac.kr}\\

{\it Topology and Geometry Research Center, Kyungpook National University,\\
Taegu 702-701, Korea}\\

\vspace{10mm}

\end{center}

\begin{center}
{\bf ABSTRACT}
\end{center}
In this paper, we study the symmetry of quantum torus with the
concept of crossed product algebra. As a classical counterpart, we
consider the orbifold of classical torus with complex structure
and investigate the transformation property of classical theta
function. An invariant function under the group action is
constructed as a variant of the classical theta function. Then our
main issue, the crossed product algebra representation of quantum
torus with complex structure under the symplectic group is
analyzed as a quantum version of orbifolding.
 We perform this analysis with Manin's so-called model II quantum theta function
 approach.
The symplectic group $Sp(2n,{\Z})$ satisfies the consistency
condition of crossed product algebra representation. However, only
a subgroup of $Sp(2n,{\Z})$ satisfies the consistency condition
for orbifolding of quantum torus.
\\


\noindent

\thispagestyle{empty}

\newpage
\section*{1. Introduction}

Classical theta functions \cite{mumford} can be regarded as state
functions on classical tori, and have played an important role in
the string loop calculation \cite{jp,gsw}.
Recently, Manin \cite{manin1,manin2,manin3} introduced the concept
of quantum theta function as a quantum counterpart of classical
theta function. In our previous work \cite{ek1}, we clarified the
relationship between Manin's quantum theta function and the theta
vector \cite{schwarz01,ds02,kl03} which Schwarz  introduced
earlier. In \cite{ek1}, we showed the connection between the
classical theta function and the so-called $kq$ representation
which appeared in the physics literature \cite{zak72,bgz75}. Then
we showed that the Manin's quantum theta function corresponds to
the quantum version of the $kq$ representation.
 In the
physics literature, quantum theta functions are related with
noncommutative solitons \cite{mm01} whose solutions are given in
terms of projection operators \cite{gms00,mm01,ghs01}.
Under the lattice translation, quantum theta function maintains
the transformation property of classical theta function. Manin's
construction of quantum theta function \cite{manin2,manin3} is
based on the algebra valued inner product of the theta vector, and
this construction is a generalization of
 Boca's  construction of projection operators on the
${\Z}_4$-orbifold of noncommutative two torus \cite{boca99}.

In the algebra valued inner product one can make the inner product
of the dual algebra, the representation of the perpendicular
lattice space, be invertible or proportional to the identity
operator. This makes the algebra valued inner product be a
projection operator \cite{rief88}.
In Boca's work \cite{boca99}, the projection operators on the
${\Z}_4$-orbifold of noncommutative two torus were constructed
based on the algebra valued inner product that Rieffel
\cite{rief88} used in his classic work on projective modules over
noncommutative tori.

One can consider a symmetry group defining an orbifold from the
view point of the crossed product algebra of the original algebra
with the given symmetry group \cite{comb,ks9900,mm01}. Therefore
in order to find a representation of an orbifold algebra, one has
to find a representation of the group compatible to that of the
original algebra. In the Boca's work, the action of
${\Z}_4$-quotient was represented as the Fourier transformation,
and the algebra valued inner product was evaluated with the
eigenstates of Fourier transformation \cite{boca99}.

When the consistency conditions for the representation of crossed
product algebra are fulfilled, the group of the crossed product
algebra behaves as a symmetry group of the original algebra. The
consistency conditions for crossed product algebra are basically
having compatible actions of the group acting on the original
algebra and on the module.

 For quantum tori,
there are two types of symmetries. One is a symmetry under a group
action, and the other is a symmetry under deformation of the
algebra, the so-called Morita equivalence \cite{schwarz98}.
Here, we restrict our discussion to the symmetry under a group
action that is not related to the Morita equivalence.

In this paper, we first consider classical functions under
orbifolding of torus and try to find an invariant function under
the symplectic group $Sp(2n,{\Z})$. We then look into the
representation of crossed product algebra as a way of orbifolding
in the quantum (noncommutative) case.


The organization of the paper is as follows.
 In section 2, we review orbifolding of classical torus and construct
 an invariant function under the
 action of $Sp(2n, {\Z})$.
 In section 3, we first review the crossed product algebra and its
 consistency conditions.
Then, we check the consistency conditions of our crossed product
algebra with the group $Sp(2n, {\Z})$ via the approach of Manin's
model II quantum theta function.
 In section 4, we conclude with discussion.
\\




\section*{2. Orbifolding and classical theta function}\label{orb-ctf}

In this section, we first consider orbifolding under a group
action.
A classical function $f$ on an orbifold $X=M/G$ should satisfy
\begin{align}
f(g \cdot x)=f(x), ~~{}^\forall g \in G, ~~ x \in M . \label{obf1}
\end{align}

Now, we consider the case in which $M$ is a complex torus. Let
$M=\C^n/\Lambda$ ($\Lambda \cong \Z^{2n})$ be a complex torus. If
$M$ can be embedded in a projective space $\C P^N$ for some $N$,
then it is called an abelian variety.
 For $M$ to be an abelian
variety, there must exist a polarization, a positive line bundle
on $M$. A positive line bundle $L$ on $M$ should satisfy that
$\int_C c_1(L)>0$, for any curve $C$ in $M$, where $c_1(L)$ is the
first Chern class of $L$ as an element of $ H^2(M,\Z) \cap
H^{1,1}(M,\R)$. 
Explicitly, $c_1(L)=\sum \delta_\alpha dx_\alpha \wedge dy_\alpha
= \sum q_\beta dz_\beta \wedge d\bar{z}_\beta$, $\delta_\alpha \in
\Z$, and $q_\beta$ is pure imaginary.
In particular, if $\delta_\alpha=1$, for all $\alpha$, then the
abelian variety is called {\it principally polarized}. The moduli
space $\mathfrak{M}$ of principally polarized abelian varieties is
the collection of the pair
  \{ $ (M,L) | M=\C^n/\Lambda $, $L$ is a
principally polarized line bundle \} .
Let $\mathbb{H}_n=\{T | T \in M_n(\C), T^t= T, Im T >0\}$ on which
$Sp(2n, \Z)$ acts as follows:
\[
g \cdot T = (AT + B)(CT+D)^{-1}, ~~ {\rm for} ~~ g = \left(
\begin{array}{cc} A&B\\C&D\end{array} \right)  \in
Sp(2n,{\Z}).
\]
%
%
%
%
Then, $\mathfrak{M}=Sp(2n,\Z) \backslash \mathbb{H}_n$.

Now, we consider an action of a group $G$ on $M$. In other words,
a map from $G \times M$ to $M$, such that for every $g \in G$, $g$
is an automorphism of $M$ preserving complex structure $T$ and the
group structure. Then, $g$ induces a linear map from $\C^n$ to
$\C^n$ sending $\Lambda$ to $\Lambda$.
It means that $g$ belongs to $GL(n, \C )$ and also $GL(2n, \Z)$
which is given in terms of the basis of $\Lambda ~ (\cong
\Z^{2n})$, whose determinant  is $\pm 1$. Additionally, if we
impose that $g$ preserves $L$, then $g$ preserves $c_1(L)$, so
that
\[
c_1(L) = \sum dx_\alpha \wedge dy_\alpha
     =g^* (c_1(L))= \sum d(g^* x_\alpha) \wedge d ( g^* y_\alpha ).
\]
It implies that $ g \in Sp(2n,\Z)$. Then we can define an orbifold
$M/G$ with the preserved polarization $L$.

If $g \in GL(n, \C )$ and $ g \in Sp(2n,\Z)$, then $T' = g \cdot T
= T$ as we see below. Hence, only a subgroup of $Sp(2n,{\Z})$,
namely $ GL(n, \C ) \bigcap Sp(2n,{\Z})$ , acts as a symmetry
group for orbifolding.
\\
 For $ g \in Sp(2n,\Z)$, it acts on the basis as follows:
\[
 \left(
\begin{array}{cc} A&B\\C&D\end{array} \right) \left(
\begin{array}{c} T \\ I \end{array} \right) =
\left(
\begin{array}{c} AT + B \\ CT + D \end{array} \right)
\sim \left(
\begin{array}{c}
(AT + B) (CT + D)^{-1} \\ I \end{array} \right) = \left(
\begin{array}{c} T' \\ I \end{array} \right).
 \]
On the other hand, for $g \in GL(n, \C )$ it acts as follows:
\[
 \left(
\begin{array}{c} T \\ I \end{array} \right) \cdot g^{t} =
\left(
\begin{array}{c} T \cdot g^{t} \\ I \cdot g^{t} \end{array} \right)
\sim \left(
\begin{array}{c}
T \cdot g^{t} \cdot g^{-t} \\ I \end{array} \right) = \left(
\begin{array}{c} T \\ I \end{array} \right).
 \]
Since the two actions should yield the same result, we get to the
result that $T' = g \cdot T = T$.



We now consider whether the classical theta function $\theta$ is
well defined on the above mentioned orbifold. The classical theta
function $\theta$ is a complex valued function on ${\C}^n$
satisfying the following relation.
\begin{align}
\theta(z+\lambda')& =\theta(z) ~~~~~ {\rm for} ~~~ z \in {\C}^n ,
~ \lambda' \in \Lambda', \label{ct1} \\
\theta(z+\lambda) & = c(\lambda) e^{q(\lambda , z )} \theta (z)
~~~~~ {\rm for} ~~~ \lambda \in \Lambda , \label{ct2}
\end{align}
where $~ \Lambda' \bigoplus \Lambda \subset {\C}^n ~$ is a
discrete sublattice of rank $2n$ split into the sum of two
sublattices of rank $n$, isomorphic to $~{\Z}^n~$, and $~c:~
\Lambda \rightarrow {\C}~ $ is a map and $~q: ~ \Lambda \times
{\C} \rightarrow {\C}~$ is a biadditive pairing linear in $z$.


The above property reflects the fact that the classical theta
function lives on ${\C}^n$ not on ${\T}^{2n}$.
The function $\theta(z,T)$ satisfying (\ref{ct1}) and (\ref{ct2})
can be defined as
\begin{align}
\theta(z,T) = \sum_{k\in {\Z}^n} e^{ \pi i (k^t T k + 2 k^t z)}
\label{ct3}
\end{align}
where $T \in \mathbb{H}_n $.
 With the above definition, $c(\lambda)$ and
$q(\lambda , z )$ in (\ref{ct2}) are given explicitly by
$c(\lambda)=e^{- \pi i m^t T m  }$ and $q(\lambda , z )=- 2\pi i
m^t z ~ $ when $ \lambda = T m, ~ m \in {\Z}^n$.
Also
 $z \in {\C}^n$ transforms as
\begin{align}
g \cdot z=z' = (CT+D)^{-t}z, ~~ {\rm for} ~~ g = \left(
\begin{array}{cc} A&B\\C&D\end{array} \right)  \in
Sp(2n,{\Z}), \label{modtrs2}
\end{align}
where ``$-t$" denotes the transposed inverse.
Under this modular transformation, the classical theta function
transforms as follows.
\begin{align}
g \cdot \theta(z,T)= \theta(z',T') = \xi_g \det (CT +
D)^{\frac{1}{2}} e^{\pi i \{z^t (CT+D)^{-1} Cz \} } \theta(z,T) ,
~~ {}^\forall g \in Sp(2n, {\Z}) \label{modtrs3}
\end{align}
where $\xi_g$ is an eighth root of unity depending on the group
element $g$ \cite{mumford}.
Now, we like to find a compatible function on the orbifold in
which the complex structure is preserved, $g \cdot T = T$.
 For this, we first try to construct a new function which has the
symmetry properties of the classical theta function, (\ref{ct1})
and (\ref{ct2}).
We define a new function as a linear combination of the classical
theta functions under the group action:
\begin{align}
\Theta_1(z,T) = \sum_{g \in G} g \cdot \theta(z,T) .
\label{newth1}
\end{align}
Clearly the above function is invariant under the group action,
\begin{align}
h \cdot \Theta_1(z,T) = \sum_{g \in G} h \cdot g \cdot
\theta(z,T)= \sum_{g' \in G}
 g' \cdot \theta(z,T) = \Theta_1(z,T), ~~ {}^\forall h \in G .  \label{newthtrs1}
\end{align}
However, this function does not possess the symmetry properties of
the classical theta function (\ref{ct1}) and (\ref{ct2}).
This is because the condition (\ref{ct1}) is not satisfied by
$\Theta_1(z,T)$, since
\begin{align}
 g \cdot \theta(z + \lambda',T)=  \theta(g \cdot (z+\lambda'),g \cdot T)
=  \theta(g \cdot z+  g \cdot \lambda', T)
  \neq \theta(g \cdot z,T),
 \label{gptrs1}
\end{align}
where $g \cdot \lambda' \in \Lambda + \Lambda'$ for some $\lambda'
\in \Lambda'$ due to the modular transformation $g \cdot \lambda'=
(CT +D)^{-t} \lambda' $.
 For the condition (\ref{ct2}), each $g \cdot \theta$ in $\Theta_1(z,T)$ in (\ref{newth1})
 gets a different factor for a
lattice shift in $\Lambda $:
\begin{eqnarray}
 g \cdot \theta(z + \lambda,T) & = &  \theta(g \cdot (z+\lambda),g \cdot T)
    =   \theta(g \cdot z+ g \cdot  \lambda,  T)
  \nonumber \\
 & \neq & \theta(g \cdot z + \lambda,T) ~~{\rm for} ~~ \lambda \in
  \Lambda ,
 \label{gptrs2}
\end{eqnarray}
since again $g \cdot \lambda = (CT +D)^{-t} \lambda  \neq \lambda$
and belongs to $ \Lambda + \Lambda'$ in general. Thus the function
$\Theta_1(z,T)$ fails to preserve the transformation properties of
the classical theta function, (\ref{ct1}) and (\ref{ct2}), though
it is a well defined function on the orbifold.

In (\ref{ct3}), the above result was due to the product $k^t z$ in
the exponent. So we need to find a new combination of this type of
product under the modular transformation that preserves the
complex structure. Since a symplectic product preserves the
complex structures, we modify the classical theta function as
follows.
\begin{align}
\widetilde{\Theta}(z,T) = \sum_{\underline{k}}  \exp \left( -\pi
H_T (\underline{k},\underline{k})  + 2 \pi i ~ {\rm Im} [H_T
(\underline{k}, z)] \right)   \label{modth}
\end{align}
where
\begin{align}
H_T(s,z)\equiv s^t ({\rm Im }T )^{-1} z^* ~~ {\rm for} ~~ s,z \in
{\C}^n . \label{symprod}
\end{align}
Here, $T$ is the complex structure given before, and
$\underline{k}$ denotes the lattice point given by $\underline{k}
= T k_1 + k_2$ with $ k_1, k_2 \in {\Z}^n$, and $ z  \in {\C}^n$
is given as usual with $ z=T x_1 + x_2$ with $ x_1, x_2 \in
{\R}^n$. Here, we notice that ${\rm Im} [H_T (\underline{k}, z)]=
{\rm Im}[\underline{k}^t ({\rm Im }T )^{-1} z^*] = k_1^t x_2 -
k_2^t x_1$. 
If we denote $\underline{x}$ as $z=T x_1 + x_2 \equiv
\underline{x}$ and the same for $\underline{y}=T y_1 +y_2$ with
$y_1, y_2 \in {\R}^n$, then $H_T(\underline{x},\underline{y})=
\underline{x}^t ({\rm Im }T )^{-1} \underline{y}^*$ is an
invariant combination under the modular transformation,
$T'=(AT+B)(CT+D)^{-1}, ~ \underline{x}' = (CT+D)^{-t}
\underline{x}$ and the same for $\underline{y}$, for any $ \left(
\begin{array}{cc} A&B\\C&D\end{array} \right)  \in
Sp(2n,{\Z})$. One can check that the above transformation of the
complex coordinate $\underline{x}$ is compatible with the
following coordinate transformation in the real basis.
\begin{align}
\left(
\begin{array}{c} x_1'\\x_2'\end{array} \right) = \left(
\begin{array}{cc} A&B\\C&D\end{array} \right)^{-t} \left(
\begin{array}{c} x_1\\x_2\end{array} \right).
\label{xmodtrs}
\end{align}


The first term in the exponent in (\ref{modth}) is invariant under
the modular transformation as we shall see in the next section,
and the second term is also invariant since it is a symplectic
product that preserves the complex structure. Thus, our modified
theta function $\widetilde{\Theta}$ is invariant under the modular
transformation, and it is a well defined function on the above
orbifold.

In fact, we can view this as follows. The classical theta function
$\theta$ in (\ref{ct3}) is summed over only one of the two
${\Z}^n$ lattices $\Lambda, \Lambda'$ in the $2n$-torus. Our
modified theta function $\widetilde{\Theta}$ is summed over the
both lattices, and its property under lattice translation is
changed from that of the classical theta function. The new
function $\widetilde{\Theta}$ is invariant under the lattice
translation in both directions, $\Lambda$ and $\Lambda'$. And this
property is preserved under the group action.

In general, for a manifold $M$ on which a group $G$ is acting, one
can define invariant functions on $M$ under the action of the
group $G$ as the functions on the orbifold $M/G$. In the next
section, we will do a quantum counterpart of the above analysis
with crossed product algebra.
\\

%
\section*{3. Quantum torus with crossed product algebra}\label{tqfob}

In order to consider an orbifolding of quantum torus, we have to
express the group action in terms of the representation of the
crossed product algebra. So, in this section we first review
briefly about the crossed product algebra and its representation,
then we will investigate the representation of crossed product
algebra for orbifolding.

%
\subsection*{3.1 Crossed product algebra}\label{cpalg}




We now consider the crossed product algebra and its representation
\cite{comb,mm01}.

Let $G$, a group, act on an algebra ${\cal A}$. More explicitly there
is a group homomorphism
\[
\varepsilon: ~~  G \rightarrow  Aut ({\cal A}).
\]
Then we define the crossed product algebra
${\cal B}= {\cal A} \rtimes_\varepsilon G$, which is $ {\cal A}
[G]= \{b~|~b~:~G \rightarrow {\cal A} \}$ as a set.
And we formally express $b \in {\cal B}$ as $\sum_{g \in G} b_g
g$, where $b_g = b(g) \in {\cal A}$.
Here, addition and scalar product are defined naturally. To define
multiplication we require the following relation:
\begin{equation}
 g \cdot b_{g'} g^{-1} =  \varepsilon(g) (b_{g'}), ~~~ g, g' \in G , ~~~ b_{g'} \in {\cal A}.
 \label{consprd}
 \end{equation}
 For $b,c, d \in {\cal B}$ with $b=\sum_{g\in G} b_g g ,
~ c=\sum_{g'\in G} c_{g'}g', ~ d=\sum_{h\in G} d_{h}h $, we can
express the multiplication $b*_{\varepsilon} c =d$ as
\begin{align}
 b \ast_\varepsilon c &= \sum_g b_g g \cdot \sum_{g'} c_{g'}g' = \sum_{g,g'} b_g g \cdot c_{g'}g'
\nonumber
\\
& = \sum_{g,h} b_g \varepsilon(g) (c_{g^{-1} h}) h \nonumber
\\
 & = \sum_{h}d_{h} h  = d    , \label{crspdf}
\end{align}
where we set $g' = g^{-1} h, ~ d_{h}=b_g \varepsilon(g) (c_{g^{-1}
h}) $, and used the relation (\ref{consprd}).

 If there are representations
$\pi , ~ u$ which are a representation of ${\cal A}$ and a
representation of the group $G$, respectively,  on a module ${\cal
H}$,
\[
\pi : ~ {\cal A} \rightarrow  End ({\cal H}) , ~~~ u: ~ G
\rightarrow Aut ({\cal H}) ,
\]
then (\ref{consprd}) leads to the following condition for any
representation of the crossed product algebra should satisfy:
\begin{align}
 u(g) \pi (a) u(g^{-1})= \pi ( \varepsilon(g) (a)), ~ {}^\forall a \in {\cal
 A}, ~ {}^\forall g \in G .
 \label{covar}
\end{align}
Furthermore, if there exists an ${\cal A}$ valued inner product
${}_{\cal A} \ll  ,  \gg$ on ${\cal H}$, then the following should
be also satisfied for consistency \cite{comb},
\begin{align}
 \varepsilon(g) ({}_{\cal A} \ll \xi , \eta \gg) = {}_{\cal A} \ll u(g)\xi, u(g) \eta
 \gg , ~~ {\rm for} ~~ g \in G, ~~ \xi, \eta \in {\cal H}.
 \label{cscrp}
\end{align}
Here, $ {}_{\cal A} \ll \xi , \eta \gg   $ denotes the ${\cal
A}$-algebra valued inner product to be defined below, which
belongs to ${\cal A}$.
 We changed the notation for the algebra valued
inner product from the single bracket in our previous work
\cite{ek1} to the double bracket to distinguish it from the usual
scalar product which we will denote with the single bracket below.



%
\subsection*{3.2 Symmetry transformation }\label{tqf-ap}


In \cite{manin3}, Manin constructed the quantum theta function in
two ways which he called model I and model II. The model I basically
follows the Rieffel's way of constructing projective modules over
noncommutative tori. Thus in the model I, one deals with  Schwartz
functions on ${\R}^{n}$ for complex $n$-torus. And the scalar
product is defined as
\begin{align}
<\xi, \eta> = \int \xi(x_1) \overline{\eta(x_1)} d \mu(x_1), ~~
x_1 \in {\R}^{n} \label{scprod1}
\end{align}
where $\overline{\eta(x_1)}$ denotes the complex conjugation of
$\eta(x_1)$, and $d \mu(x_1)$ denotes the Haar measure in which
${\Z}^n$ has covolume 1.

In the model II, one deals with holomorphic functions on ${\C}^n$,
and the scalar product is defined as
\begin{align}
<\xi, \eta>_T = \int_{{\C}^n} \xi(\underline{x})
\overline{\eta(\underline{x})} e^{- \pi H_T(\underline{x},
\underline{x})} d \nu \label{scprod2}
\end{align}
where $d \nu $ is the translation invariant measure making
${\Z}^{2n}$  a lattice of covolume 1 in ${\R}^{2n}$.
Here, $\underline{x}= T x_1 + x_2 $ with $ x_1, x_2 \in {\R}^n $.
The complex structure $T$ is given by an $n \times n$ complex
valued matrix, and $H_T(\underline{x}, \underline{x})=
\underline{x}^t ({\rm Im}T)^{-1}\underline{x}^*$ as in
(\ref{symprod}).




Now, we do the analysis with the model II quantum theta function.
 For consistency of the representation of a crossed product
algebra ${\cal B}= {\cal A} \rtimes G$, we need to define the
following as explained in
the previous subsection :\\
(I) $~~\pi~ :~ {\cal A} \rightarrow End(\cal H)$
\\
(II) $~u~:~G \rightarrow Aut(\cal H)$
\\
(III) $\varepsilon~:~G \rightarrow Aut({\cal A})$, such that
 $u(g)\pi(a)u(g^{-1}) = \pi (\varepsilon(g)(a))$
\\
(IV) $\ll,\gg ~: ~ {\cal H} \times {\cal H} \rightarrow {\cal A}$
, such that
 $\varepsilon(g) \ll f,h\gg =\ll u(g)f , u(g)h\gg$.
%
%

 Let $M$ be any
locally compact Abelian group
 and $\widehat{M}$ be its dual group and define ${\cal G} \equiv M
\times \widehat{M}$. And, let $\pi$ be a representation of ${\cal
G}$ on $L^2(M)$  such that
\begin{align}
\pi_x \pi_y = \alpha (x,y) \pi_{x+y} =\alpha (x,y)
\overline{\alpha}(y,x) \pi_y \pi_x ~~~ {\rm for}~~ x,y \in {\cal
G} \label{ccl}
\end{align}
where $\alpha$ is a map $ \alpha : ~ {\cal G} \times {\cal G}
\rightarrow {\C}^* $ satisfying
\[ \alpha(x,y)
=\alpha(y,x)^{-1} , ~~~ \alpha(x_1 + x_2 , y) = \alpha(x_1 , y)
\alpha (x_2 , y).  \]
%
We also define $S(D)$ as the space of Schwartz functions on $D$
which we take as a discrete subgroup of ${\cal G}$.
 For $\Phi \in S(D)$, it can be expressed as $\Phi = \sum_{w \in D} \Phi(w) e_{D,
 \alpha}(w)$ where $e_{D, \alpha}(w)$ is a delta function with
 support at $w$ and obeys the following relation.
\begin{equation}
e_{D, \alpha} (w_1) e_{D, \alpha} (w_2) = \alpha(w_1,w_2) e_{D,
\alpha} (w_1 +w_2) . \label{ccld}
\end{equation}


 From now on, we take $M$ as $\R^n $.
Let ${\cal A}$ be $S(D)$ valued functions on $\mathbb{H}_n$. More
explicitly
\begin{align}
{\cal A} = S(D) \otimes {\cal F}(\mathbb{H}_n)= \{a~|~a :
\mathbb{H}_n \rightarrow S(D) \},
\end{align}
where ${\cal F}(\mathbb{H} _n)$ is an algebra of smooth complex
functions on $\mathbb{H}_n$.
Then $a(T)=\sum_{w \in D} a_{T,w}e(w)$, where $a_{T,w} \in \C$.
Let ${\cal H}$ be given as follows.
\begin{align}
{\cal H} = \{f~ &|~f: \R^n \times \widehat{\R}^n \times
\mathbb{H}_n \rightarrow \C, \nonumber
\\
& <f(x,T),f(x,T)>_T = \int|f(x,T)|^2 e^{-\pi H_T(x,x)} dx <
\infty,~ {}^\forall T \}
\end{align}
where $x \in \R^n \times \widehat{\R}^n, ~~ T \in \mathbb{H}_n$
and from here on $H_T(x,y)$ that we used above denotes $H_T
(\underline{x},\underline{y})$ defined in the section 2 for
notational convenience.
%
 In other words, $\cal H$ are global sections of $\mathbb{H}$,
a vector bundle over $\mathbb{H}_n$, where the fiber over $T$ is
\begin{align}
{\mathbb H}_T =\{ \xi~|~\xi :\R^n \times {\widehat \R}^n
\rightarrow \C , ~~<\xi,\xi>_T \ < \infty \}.
\end{align}

Let the group $G$ be $Sp(2n,\Z)$ and we now carry out the steps
(I) through (IV) that we listed above.
\\

\noindent (I) Before we define $\pi$, we need to define a map
$\pi_0$ from $S(D)$ to
 $End(\cal H)$:
 \[ \pi_0 : e(w) \rightarrow  \pi_w  ~~~ {\rm for} ~~ w \in D \]
where
\begin{align}
(\pi_w f)(x,T)= e^{-\pi H_T (x,w)-\frac{\pi}{2}H_T (w,w)}f(x+w,T).
\end{align}
Let $a \in {\cal A}$, where $a(T)=\sum_w a_{T,w} e(w)$. Now, we
define $\pi$ as follows.
\begin{align}
(\pi(a)f)(x,T)=[\pi_0(a(T))f](x,T).
\end{align}

\noindent (II) We define $u$ as follows.
\begin{align}
(u(g)f)(x,T)=f(g \cdot x, g \cdot T),
\end{align}
where $g = \left( \begin{array} {cc} A&B\\C&D
\end{array} \right) \in Sp(2n,\Z)$, ~
$g \cdot x = \left(
\begin{array} {cc} A&B\\C&D
\end{array} \right)^{-t}x$,  and $g \cdot T = (AT+B)(CT+D)^{-1}$.

 For the remaining steps we need to use the following two lemmas. \\
Lemma 1 :
\begin{align}
H_T(x,y) = H_{g \cdot T}(g \cdot x, g \cdot y).
 \label{lm1}
 \end{align}
Lemma 2 :
\begin{align}
<f,h>_{g \cdot T} = <u(g)f,u(g)h>_T.
\label{lm2}
\end{align}

\noindent
Proof of the lemma 1:\\
We first want to show that
\begin{align}
Im(g \cdot T)=Im((AT+B)(CT+D)^{-1})=(C\overline{T} +D)^{-t}
Im(T)(CT+D)^{-1}.
 \label{corl}
\end{align}
Then the proof of the lemma 1 is given by the following steps.
\begin{align*}
H_{g \cdot T}(g \cdot x,g \cdot y)&=((CT+D)^{-t}
\underline{x})^t(Im(g \cdot T))^{-1}((CT+D)^{-t} \underline{y})^*
\\
&= \underline{x}^t (CT+D)^{-1}(CT+D)(Im(T))^{-1}(C
\overline{T}+D)^t (C\overline{T}+D)^{-t} \underline{y}^*
\\
&= \underline{x}^t (Im(T))^{-1} \underline{y}^* = H_T(x,y).
\end{align*}
Thus, we only have to show (\ref{corl}). We can prove it with the
three generators of $Sp(2n,\Z)$ \cite{mumford}.
\begin{align}
 i)~~ g &= \left( \begin{array} {cc} A&0\\0&A^{-t} \end{array}
\right),~~~A \in GL(n,\Z)
\\
ii)~~ g &= \left( \begin{array} {cc} I&B\\0&I \end{array}
\right),~~~B^t=B~,~~B \in gl(n,\Z)
\\
iii)~~ g &= \left( \begin{array} {cc} 0&-I\\I&0 \end{array}
\right).
\end{align}
 For the first two cases, (\ref{corl}) can be shown trivially. For the
case iii), we need to show the following:
\begin{align}
Im~T' = \overline{T}^{-t} (Im~T)T^{-1}= \overline{T}^{-1}
(Im~T)T^{-1} \label{auxl}
\end{align}
where $T' = g \cdot T = - T^{-1}$.
\\
Now, we prove (\ref{auxl}).
\\
Let $T=T_1 +iT_2$ and $T' = T_1' + i T_2'.$ Then from $T'T =-I$,
we get $T_1'T_1-T_2'T_2=-I$ and $T_2'T_1+T_1'T_2=0$.
Then the statement we want to prove becomes
$T_2'=\overline{T}^{-1}T_2 T^{-1}$, or equivalently,
\begin{align}
\overline{T}T_2' T =T_2.
\label{crf}
\end{align}
The left hand side of (\ref{crf}) is
\begin{align}
L.H.S. &= (T_1-iT_2)T_2'(T_1+iT_2)
\nonumber \\
&= (T_1 T_2' T_1 +T_2 T_2' T_2) + i(-T_2 T_2' T_1 +T_1 T_2' T_2) .
\nonumber
\end{align}
Using $T_1'T_1-T_2'T_2=-I$ and $T_2'T_1+T_1'T_2=0$  together with
the property that $T_i, T_i'$ are symmetric, then we can easily
show that
$$L.H.S. = T_2=R.H.S.$$


\noindent
Proof of the lemma 2: \\
The left hand side of (\ref{lm2}) is
\begin{align*}
L.H.S. &= <f,h>_{g \cdot T}
\\
&= \int f(x,g \cdot T) \overline{h(x,g \cdot T)} e^{-\pi H_{g
\cdot T}(x,x)} dx ,
\end{align*}
and the right hand side of (\ref{lm2}) is
\begin{align*}
R.H.S. &=<u(g)f,u(g)h>_T
\\
&= \int f(g \cdot x,g \cdot T) \overline{h(g \cdot x,g \cdot T)}
e^{-\pi H_T(x,x)} dx
\\
&= \int f(x,g \cdot T) \overline{h(x,g \cdot T)} e^{-\pi H_{g
\cdot T}(x,x)} dx,
\end{align*}
where we used the lemma 1 in the final step.
\\

\noindent
(III) We define $\varepsilon : G \rightarrow
Aut(\cal{A})$ such that $u(g)\pi(a) u(g^{-1})=\pi(\varepsilon
(g)(a))$.
\\Let $a(T)$ be $\sum a_{T,w} e(w)$.
The left hand side can be evaluated as follows.
\begin{align*}
(u(g)\pi(a)u(g^{-1})f)(x,T) &= (\pi(a) u(g^{-1})f)(g \cdot x,g
\cdot T)
\\
&= \sum_w a_{g \cdot T,w} e^{-\pi H_{g \cdot T}(g \cdot
x,w)-\frac{\pi}{2} H_{g\cdot T}(w,w)} f(x+g^{-1} \cdot w,T)
\end{align*}
If we define $\varepsilon(g)(a)(T)=\sum_w a_{g\cdot
T,w}e(g^{-1}\cdot w)$, then the right hand side is given by 
\begin{align*}
\pi(\varepsilon(g)(a)f)(x,T) &= \sum_w a_{g \cdot T,w} \pi
(g^{-1}\cdot w)f(x,T)
\\
&= \sum_w a_{g \cdot T,w} e^{-\pi H_{g \cdot T}(g \cdot
x,w)-\frac{\pi}{2} H_{g \cdot T}(w,w)} f(x+g^{-1} \cdot w,T).
\end{align*}
In the last equality we used the lemma 1. So those two sides are
equal. Using the lemma 1, one can also show the following.
\begin{align}
u(g)\pi_w u(g^{-1})=\varepsilon(g)\pi_w=\pi_{g^{-1}\cdot w}.
\label{auxl2}
\end{align}


\noindent
(IV) We define an $\cal{A}$-valued  inner product on
$\cal H$ as follows.
\begin{align}
\ll f,h \gg (T)=\sum_w <f,\pi_w h>_T e(w)
\end{align}
where $<f,\pi_w(h)>_T=<f(x,T),\pi_w h(x,T)>_T$.
\\
In other words if $a=\ll f, h \gg$ then $a_{T,w}=<f,\pi_w h >_T$.\\
Now, we want to check that $\varepsilon(g)\ll f,h \gg = \ll
u(g)f,u(g)h \gg$ holds.
\\ Recall that
\[
 \varepsilon(g)(a)(T) = \sum_{w} a_{g \cdot T, w} e(g^{-1} \cdot
 w).
\]
The left hand side is given by
\begin{align*}
 (\varepsilon(g) (\ll f,h \gg))(T) &=\sum_w <f, \pi_w h>_{g \cdot T} e(g^{-1} \cdot w)
\\
&=\sum_w <f, \pi_{g \cdot w} h>_{g \cdot T} e(w)).
\end{align*}
The right hand side is given by
\begin{align*}
\ll u(g)f, u(g)h \gg_T &= \sum_w <u(g)f, \pi_w u(g)h>_T e(w)
\\
&= \sum_w <f, u(g)^{-1} \pi_w u(g)h>_{g \cdot T} e(w)
\\
&= \sum_w <f, \pi_{g\cdot w}h>_{g \cdot T} e(w),
\end{align*}
where we used the lemma 2 and (\ref{auxl2}).
%

\subsection*{3.3 Orbifolding quantum torus}\label{orbqct}

We consider an orbifolding of quantum torus with a polarized
complex structure $T$.
The symmetry group preserving the polarized
complex structure is the subgroup $G_T=\{g\in Sp(2n,\Z) | g \cdot
T=T \}$ of $Sp(2n,\Z)$. 
Orbifolding the quantum torus with a complex structure $T$
corresponds  to the crossed product algebra discussed in the
previous section with fixed $T$.

Let $A_T=S(D)$ and $\mathbb{H}_T=\{f_T|f_T : \R^n \times \R^n \to
\C , \lVert f_T \rVert^2= \int |f_T(x)|^2 e^{-\pi H_T(x,x)} dx <
\infty \}.$
Now, we can define the crossed product algebra, $A_T \rtimes G_T$,
naturally from the construction in the section 3.2:
\begin{enumerate}
\item $\pi_T$ : $A_T \to End(\mathbb{H}_T)$
\item $u_T$ : $G_T \to Aut(\mathbb{H}_T)$
\item $\varepsilon_T$ : $G_T \to Aut(A_T)$ such
that $u_T(g)\pi_T(a)u_T(g^{-1})=\pi_T(\varepsilon_T(g)(a))$
\item
$\ll,\gg_T$ : $\mathbb{H}_T \times \mathbb{H}_T \to A_T$ such that
$\varepsilon_T(g) \ll f_T,h_T \gg _T =\ll u_T(g)f_T, u_T(g)h_T \gg
_T$.
\end{enumerate}
Here, $\pi_T, ~ u_T, ~ \varepsilon_T, ~ \ll,\gg_T, ~ f_T$ satisfy
the following relations:
\begin{align*}
(\pi_T (a(T))f_T)(x) & =  (\pi(a)f)(x,T),
\\
(u_T(g)f_T)(x) & =  (u(g)f)(x, T),
\\
( \varepsilon(g)(a))(T) &  =  \varepsilon_T(g) (a(T)),
\\
\ll f_T, h_T \gg_T  & =  \ll  f , h \gg (T),
\end{align*}
where $f_T(x)= f(x,T), ~ a  \in   S(D) \otimes  {\cal F} (
\mathbb{H}_n)$ and $g \in G_T$.
If we choose $f(x,T)=1$, then $\varepsilon(g) \ll 1, 1 \gg = \ll
u(g)1, u(g)1 \gg = \ll 1, 1 \gg $, and thus $\ll 1, 1 \gg $ which
belongs to the algebra ${\cal A}$ is $Sp(2n, \Z)$ invariant.
Since $\ll 1, 1 \gg (T) = \sum_{w \in D} e^{-\frac{\pi}{2} H_T(w,w)}
e(w) $ is the Manin's model II quantum theta function, this also
tells us that the model II quantum theta function is well defined on
the orbifolds of quantum complex torus. We further notice that
Boca's projection operator \cite{boca99} on the $ \Z / 4 \Z $
orbifold of quantum 2-torus with $T=i$ corresponds to a special case
of this construction.
\\



\section*{4. Conclusion }

In this paper, we investigate the symmetry of quantum torus with
the group $Sp(2n, {\Z})$.

%
 First, we investigate the orbifolding of classical complex torus.
It turns out that the orbifold group for complex n-torus leaving
the complex structure and its polarization intact is a subgroup of
 $Sp(2n, {\Z})$.
Also, the classical theta function is not invariant under the
$Sp(2n, {\Z})$ transformation, and we construct a variant of the
classical theta function as an invariant function under the
transformation of $Sp(2n, {\Z})$.
Then as a quantum counterpart, we investigate the representation
of crossed product algebra of quatum torus with $Sp(2n, {\Z})$ via
Manin's model II quantum theta function approach.


In the Manin's model I approach, the dimension of the Hilbert
space variable $x_1$, which is $n$ for quantum ${\T}^{2n}$, does
not match the dimension of the fundamental representation of
$Sp(2n, {\Z})$, which is $2n$.
On the other hand, in the model II case the dimension of the
Hilbert space variable $x=(x_1,x_2)$ exactly matches that of the
group.
Therefore in the model I case the group action cannot act directly
on the variables of the Hilbert space.
Thus one has to devise a transformation such as Fourier
transformation as in the Boca's work \cite{boca99}, where
 ${\Z}_4$ acts directly
on the functions as a Fourier transformation, not on the variables
of the functions.
This type of difficulty comes from the fact that in the model I
case the number of variables of the functions is half of that of
the phase space as it is typical in the conventional quantization.
In the model II approach, the above mentioned difficulty does not
arise. The group action can be defined nicely on the module as it
acts on the variables.

In conclusion, in the model II case  $Sp(2n,{\Z})$ turns out to be
the symmetry group for the quantum torus times $\mathbb{H}_n$.
The orbifolding of quantum torus with complex structure
corresponds to the crossed product algebra, $S(D) \rtimes G_T $,
where $G_T$ is the subgroup of $Sp(2n, \Z)$ fixing the complex
structure, $g \cdot T = T$ for $g \in Sp(2n, \Z)$.
And Manin's model II quantum theta function turns out to be a well
defined function over the above orbifold of quantum torus.
\\



\vspace{5mm}

\noindent
{\Large \bf Acknowledgments}

\vspace{5mm} \noindent Most part of the work was done during
authors' visit to KIAS. The authors would like to thank KIAS for
kind hospitality. This work was supported by KOSEF
Interdisciplinary Research Grant No. R01-2000-000-00022-0.
\\


\vspace{5mm}


\end{document}